\documentclass[preprint]{emulateapj}

\usepackage{graphicx}
\usepackage{amsmath}
\usepackage{aas_macros}

\newcommand\beq{\begin{equation}}
\newcommand\eeq{\end{equation}}
\newcommand\beqar{\begin{eqnarray}}
\newcommand\eeqar{\end{eqnarray}}

\submitted{Received 2009 December 9; accepted 2010 August 27}

\shorttitle{Detectability with the anisotropy energy spectrum}
\shortauthors{Hensley, Siegal-Gaskins, and Pavlidou}

\defcitealias{NT06}{NT06}

\begin{document}

\title{The detectability of dark matter annihilation with \emph{Fermi}\\ using the anisotropy energy spectrum of the gamma-ray background}

\author{Brandon S. Hensley\altaffilmark{1,3}, Jennifer M. Siegal-Gaskins\altaffilmark{2}, and Vasiliki Pavlidou\altaffilmark{1,4}}
\email{bhensley@astro.princeton.edu}
\altaffiltext{1}{Department of Astronomy, California Institute of Technology, Pasadena, CA 91125, USA}
\altaffiltext{2}{Center for Cosmology and Astro-Particle Physics, The Ohio State University, Columbus, OH 43210, USA}
\altaffiltext{3}{Alain Porter Memorial SURF Fellow}
\altaffiltext{4}{Einstein Fellow}
\altaffiltext{5}{Current address: Peyton Hall, 4 Ivy Lane, Princeton, NJ 08544, USA}

\begin{abstract}
 The energy dependence of the anisotropy (the anisotropy energy spectrum) of the large-scale diffuse gamma-ray background can reveal the presence of multiple source populations.  Annihilating dark matter in the substructure of the Milky Way halo could give rise to a modulation in the anisotropy energy spectrum of the diffuse gamma-ray emission measured by \emph{Fermi}, enabling the detection of a dark matter signal.  We determine the detectability of a dark-matter-induced modulation for scenarios in which unresolved blazars are the primary contributor to the measured emission above $\sim 1$ GeV and find that in some scenarios pair-annihilation cross sections of order the value expected for thermal relic dark matter can produce a detectable feature.  We anticipate that the sensitivity of this technique to specific dark matter models could be improved by tailored likelihood analysis methods. 
\end{abstract}

\keywords{dark matter, galaxies: active, Galaxy: structure, gamma rays: diffuse background}

\section{Introduction}
The matter density of the universe is dominated by non-luminous, non-baryonic dark matter, though its nature remains one of the greatest outstanding questions in physics \citep[see, e.g.,][for reviews]{Bergstrom2000,Bertone2005}. 
Many weak-scale candidate dark matter particles can pair-annihilate to produce gamma-ray photons and other standard model particles, providing a means of indirectly detecting dark matter and probing its properties. The \emph{Fermi Gamma-ray Space Telescope} \citep[\emph{Fermi};][]{Atwood2009}, which is currently monitoring the gamma-ray sky from 20 MeV to more than 300 GeV, has generated intense interest in the possible detection of a dark matter annihilation signal. 

Primary targets for indirect dark matter searches in gamma rays include the Milky Way halo \citep[e.g.,][]{springel_white_frenk_etal_08, baltz_berenji_bertone_etal_08, pieri_lavalle_bertone_etal_09}, its substructure \citep[e.g.,][]{pieri_bertone_branchini_08, kuhlen_diemand_madau_08, baltz_berenji_bertone_etal_08, pieri_lavalle_bertone_etal_09}, the Galactic Center \citep[e.g.,][]{berezinsky_bottino_mignola_94, dodelson_hooper_serpico_08, serpico_zaharijas_08, baltz_berenji_bertone_etal_08}, and nearby Milky Way satellite galaxies \citep[e.g.,][]{baltz_briot_salati_etal_00, tasitsiomi_gaskins_olinto_04, strigari_koushiappas_bullock_etal_08, baltz_berenji_bertone_etal_08, abdo_others_10dwarfs}. Numerical simulations of structure formation predict that the dark matter halo of the Milky Way is populated by an enormous number of dense subhalos \citep{diemand_kuhlen_madau_etal_08,springel_white_frenk_etal_08}, and theoretical calculations indicate that cold dark matter 
could
form structures on scales of $\sim$ $M_{\oplus}$ or smaller, far below those that can currently be resolved in simulations \citep{green_hofmann_schwarz_05,profumo_sigurdson_kamionkowski_06,bringmann_09}. Outside of the Galactic center region, most photons from Galactic dark matter annihilation are expected to originate in substructure.  Almost all subhalos are likely to be too faint to be resolved individually by \emph{Fermi}, and instead will contribute to the measured diffuse emission. The radial number density distribution of the substructure is not very centrally concentrated, since subhalos near the Galactic center are easily destroyed by tidal effects. Consequently, on large angular scales, the gamma-ray flux from substructure appears almost isotropic \citep{kuhlen_diemand_madau_08, siegal-gaskins_08, springel_white_frenk_etal_08}.  On smaller angular scales, the clumpiness of the Galactic dark matter distribution would induce anisotropies \citep{siegal-gaskins_08, lee_ando_kamionkowski_09, fornasa_pieri_bertone_etal_09, Ando2009}.

We explore the possibility of detecting an annihilation signature from the substructure of the Milky Way dark matter halo in the large-scale isotropic gamma-ray background (IGRB\@).  In this work, we use the IGRB to refer to diffuse emission of extragalactic and possibly Galactic origin which appears isotropic on large angular scales, but may contain fluctuations on small angular scales.  The IGRB is thought to be composed largely of gamma rays from unresolved blazars \citep[active galaxies with jets aligned with the line of sight;][]{stecker-1996-464,NT06,Dermer07,Venters2009} and photons produced from the interaction between cosmic rays and diffuse gas in normal galaxies \citep{Pavlidou2002, TT07, ando_pavlidou_09, FPP10} with possible contributions from other sources such as Galactic and extragalactic dark matter annihilation \citep{ullio_bergstrom_edsjo_etal_02, elsasser_mannheim_05, oda_totani_nagashima_05, zavala_springel_boylan-kolchin_09, abazajian_agrawal_chacko_etal_10, abdo_others_10cosmowimps}. 

To extract information about dark matter from the IGRB, it is necessary to disentangle the dark matter annihilation signal from gamma rays arising from other sources. One way to do this is to examine the anisotropy of the gamma-ray flux on small angular scales ($\sim$ few degrees). 
Anisotropies in the IGRB are also expected in the collective emission from unresolved members of extragalactic source classes (\citealp{ando_komatsu_06, Ando2007, miniati_koushiappas_di-matteo_07, cuoco_brandbyge_hannestad_etal_08, taoso_ando_bertone_etal_09, fornasa_pieri_bertone_etal_09, ando_pavlidou_09, zavala_springel_boylan-kolchin_09}; see also \citealp{ibarra_tran_weniger_09} for the case of decaying dark matter, and \citealp{zhang_sigl_08} for dark-matter-induced anisotropies in radio emission).  When coupled with the energy spectrum, the angular power spectrum of the IGRB could be used to identify the presence of a dark matter contribution \citep{Siegal-Gaskins2009}.

In this study, we examine the prospects for using modulations in the anisotropy energy spectrum to detect a signal from Galactic dark matter in the IGRB\@.  
 In \S Section~\ref{sec:models}, we outline the models we adopt for the spectral and anisotropy properties of blazars, which are taken to be the dominant astrophysical contributor to the IGRB, and dark matter.
In \S Section~\ref{sec:detectability}, we determine the detectability of a modulation in the anisotropy energy spectrum due to a dark matter annihilation signal, explore the sensitivity of this technique to variations in the blazar and dark matter models, and identify the regions of the dark matter parameter space accessible to \emph{Fermi}.   
We discuss the results of this detectability analysis and its implications for indirect dark matter searches with \emph{Fermi} in \S Section~\ref{sec:discussion}.

\section{The blazar and dark matter models}
\label{sec:models} 

\subsection{IGRB Observables}

We use three metrics to identify a dark matter signal in the IGRB: the intensity energy spectrum, the angular power spectrum, and the anisotropy energy spectrum.
The intensity spectrum is the differential photon intensity 
$I$ as a function of energy, in units of 
photons per area per time per solid angle per energy.
If we expand a map of intensity fluctuations $\delta I$ in units of the mean intensity $\langle I \rangle$, $\delta I=(I-\langle I \rangle)/\langle I \rangle$,
in the basis of spherical harmonics, we obtain a set of coefficients $a_{\ell m}$.  
The angular power spectrum $C_{\ell}=\langle |a_{\ell m}|^{2} \rangle$ characterizes these fluctuations as a function of angular scale by giving the variance of the $a_{\ell m}$ at each multipole $\ell$.
The anisotropy energy spectrum $C_{\ell}(E)$ is the angular power spectrum evaluated at a specific multipole $\ell$ as a function of energy $E$. 
 
In the case where the total intensity, $I_{\rm tot}$, is the sum of two uncorrelated components (e.g., an extragalactic contribution $I_{\rm EG}$ and a Galactic dark matter component $I_{\rm DM}$), the angular power spectrum of the total signal is given by 
\beqar
\label{eq:clsum}
\ C_\ell^{\mbox{\tiny{tot}}} = f_{\mbox{\tiny{EG}}}^2C_\ell^{\mbox{\tiny{EG}}} + f_{\mbox{\tiny{DM}}}^2C_\ell^{\mbox{\tiny{DM}}} 
\eeqar
where $\ f_{\mbox{\tiny{EG}}} = I_{\mbox{\tiny{EG}}}/I_{\mbox{\tiny{tot}}}$ and $f_{\mbox{\tiny{DM}}} = I_{\mbox{\tiny{DM}}}/I_{\mbox{\tiny{tot}}}$
are the fractional contributions to the total intensity from extragalactic sources and dark matter, respectively.  The energy dependence of $f_{\mbox{\tiny{EG}}}$ and $f_{\mbox{\tiny{DM}}}$ make the anisotropy energy spectrum $C_{\ell}(E)$ an energy-dependent function.

\begin{figure}[ht]
\includegraphics[width=0.47\textwidth]{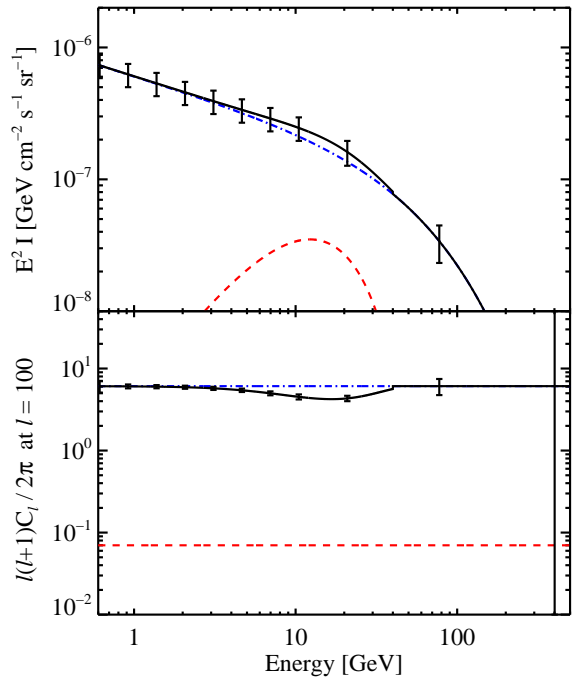}
\caption{Example intensity energy spectrum ({\it top}) and anisotropy energy spectrum ({\it bottom}) for a scenario in which the IGRB is composed of emission from Galactic dark matter substructure ({\it red dashed lines}) and unresolved blazars ({\it blue dot-dashed lines}).  The total intensity and anisotropy energy spectra are shown with solid black lines.  The proposed technique tests the measured anisotropy energy spectrum for consistency with a constant value.  See \S Section~\ref{sec:criteria} for details. \label{fig:anisoex}}
\end{figure}

Observation of a feature in the anisotropy energy spectrum of the IGRB would strongly indicate a contribution from at least two source classes.  On the other hand, if the anisotropy energy spectrum is observed to be consistent with a constant value at all energies, this would suggest that the anisotropy and intensity of the IGRB are dominated by a single source population at all energies, rather than several source populations.  An example intensity energy spectrum and corresponding anisotropy energy spectrum for a scenario in which unresolved blazars and Galactic dark matter substructure are the main contributors to the IGRB is shown in Figure~\ref{fig:anisoex}.  The dark matter contribution is subdominant in the intensity energy spectrum but produces a measurable dip in the anisotropy energy spectrum.  The error bars represent the $1\mbox{-}\sigma$ statistical uncertainties in the measurement.  The calculation of the measurement errors and the criteria used to determine the energy binning and detectability of a modulation are discussed in \S Section~\ref{sec:criteria}.

We consider a scenario in which the IGRB at energies $\gtrsim 1$ GeV is composed predominantly of emission from unresolved blazars and dark matter annihilation in Galactic substructure, as in the example shown in Figure~\ref{fig:anisoex}.  We do not include the contribution of the smooth dark matter halo, since it is negligible compared to the substructure signal at moderate to high Galactic latitudes, where the Galactic diffuse emission starts to fall off and the IGRB is most easily measured \citep[e.g.,][]{kuhlen_diemand_madau_08, Ando2009}. 
 For simplicity, we also neglect the signal from extragalactic dark matter, based on predictions that the Galactic substructure signal will exceed this emission \citep[e.g.][]{hooper_serpico_07,fornasa_pieri_bertone_etal_09}, although we note that recent work has found that by extrapolating the results of some numerical simulations the extragalactic dark matter signal can be dominant depending on the structural properties assumed for halos and subhalos \citep{pieri_lavalle_bertone_etal_09}. 

The Galactic diffuse emission is a major component of the gamma-ray sky even at high latitudes and high energies, and so we include a treatment of potential uncertainties from contamination of the IGRB by the Galactic diffuse emission.  Although we do not expect significant anisotropy in this component on $\sim1^{\circ}$ angular scales (which corresponds to a multipole $\ell\sim100$ as used in our analysis), its presence adds statistical noise, and residual contamination from this component would reduce the amplitude of the IGRB anisotropy.  We have approximated the effect of the additional noise from this component by taking $N_{\rm b}/N_{\rm s}=5$, where $N_{\rm s}$ and $N_{\rm b}$ are the number of signal and background photons, respectively, observed in the usable sky region. We also refer the reader to the recent study of \citet{Cuoco:2010jb} which considers the effect of Galactic foregrounds on the sensitivity of \emph{Fermi} to anisotropies.

In this study, we choose to focus on a multipole $\ell = 100$ as this value corresponds to an angular scale that is accessible for analysis with \emph{Fermi} data and also unlikely to be heavily contaminated by Galactic diffuse emission. We have not evaluated the optimal value of $\ell$ for this analysis, and expect that different choices of the $\ell$ at which the anisotropy energy spectrum is evaluated could increase or decrease the sensitivity of our method slightly.

\subsection{Spectral and anisotropy properties of blazars}
The intensity spectrum of the collective unresolved blazar emission at high energies is still poorly constrained both in shape and amplitude. We model the intensity energy spectrum of the blazar emission as in \citet{PavlidouVenters2007} and \citet[][their Appendix C]{pavlidou_siegal-gaskins_fields_etal_08} with the addition of a simple recipe for treating gamma-ray attenuation due to the extragalactic background light (EBL) at high energies.  The shape of the observed cumulative blazar spectrum depends on the blazar spectral index distribution, which in this formalism is  parameterized by the mean spectral index of blazars $\alpha_0$ and the spread in spectral index $\sigma_0$. Given a set of observed blazars, a likelihood analysis can be performed to find the maximum likelihood $\alpha_0$ and $\sigma_0$ as detailed in \citet{Venters2007}. Here, we derive these parameters from the sample of flat-spectrum radio quasars (FSRQs) in the \emph{Fermi}-Large Area Telescope (\emph{Fermi}-LAT) bright source list\footnote{We caution that faint gamma-ray sources with hard spectra are easier to detect than softer sources, and this might bias the derived distribution of spectral indices toward harder values if the sample used is not flux limited, which is indeed a concern with the LAT bright source list. However, especially for the brighter FSRQs that we use here, the effect is expected to be small.} \citep{Abdo2009}, since FSRQs are likely to 
dominate over BL Lac objects in their contribution to the IGRB
\citep{Dermer07}. We parameterize the normalization of the collective blazar intensity spectrum by its value at 0.26 GeV, $I_{0}=I_{\rm EG}$(0.26 GeV).  

Above a few tens of GeV, extragalactic gamma rays start suffering attenuation due to interactions with the EBL; for photons in the LAT energy range the relevant EBL photons are primarily those of ultraviolet and optical wavelengths.  EBL attenuation induces a suppression feature in the cumulative blazar spectrum, which can be calculated by integrating the luminosity function, properly accounting for the gamma-ray photon optical depth given a specific EBL model.  However, there are large uncertainties both in EBL models and the blazar luminosity function which propagate to the attenuation feature \citep{Venters2009}, so this calculation would unnecessarily complicate our blazar IGRB component model in the context of our simple analysis, without adding any robust information. Instead, we treat EBL attenuation in an approximate manner: we assume that most blazar emission is generated by FSRQs at a redshift of $z_0=1.0$ \citep[a reasonable approximation given the currently best-available luminosity function models; see, e.g.,][]{Venters2009}, and we account for attenuation using the approximate expression in \citet{Horiuchi2006}. 
We neglect the effects of cascading since we expect this to be a minor correction for the models we consider here, which have steep mean spectra (of slope $\sim 2.4$), so the amount of energy available for cascade emission is never dominant over the energy in direct emission \citep{venters_thesis_09}.

\begin{figure}[htp]
\includegraphics[width=0.47\textwidth]{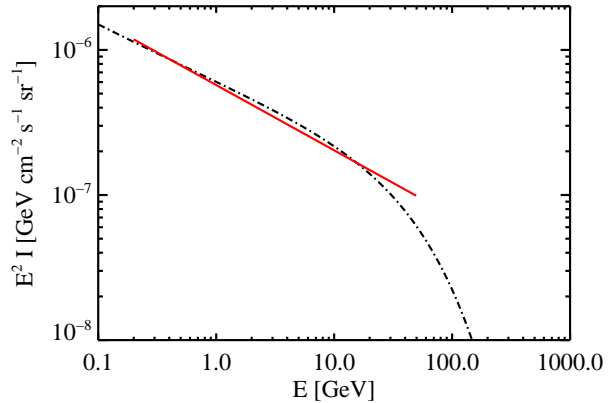}
\caption{Reference model for the blazar contribution to the IGRB ({\it black dot-dashed line}), with parameters $\alpha_{0} = 2.39$, $\sigma_{0} = 0.14$, $I_{0}=1.5 \times 10^{-5}$ GeV$^{-1}$ cm$^{-2}$ s$^{-1}$ sr$^{-1}$, and $z_0=1$. For comparison we show a power law $I \propto E^{-2.45}$ ({\it red solid line}), consistent with \emph{Fermi} measurements of the IGRB between 200 MeV and 50 GeV \citep{collaboration_10}.\label{fig:diffuse} }
\end{figure}

Our reference model for the intensity spectrum of the blazar IGRB component is shown in Figure~\ref{fig:diffuse}.
For comparison, we overplot a power law $I\propto E^{-2.45}$, consistent with \emph{Fermi} measurements of the IGRB between 200 MeV and 50 GeV \citep{collaboration_10}. The reference blazar model corresponds to $\alpha_{0} = 2.39$ and $\sigma_{0} = 0.14$, the maximum likelihood values of the spectral parameters, with a normalization at 0.26 GeV of  $I_{0}=1.5 \times 10^{-5}$ GeV$^{-1}$ cm$^{-2}$ s$^{-1}$ sr$^{-1}$ to compose the majority of the diffuse background between 0.5 and 1 GeV.  The dependence of the detectability of a dark matter feature in the anisotropy energy spectrum on the spectral parameters and intensity normalization of our blazar model is explored in \S Section~\ref{sec:blazardep}; throughout the rest of this study we adopt the reference blazar model defined here.

Here we focus on the anisotropy energy spectrum at $\ell=100$, and therefore $C_{\ell=100}^{\rm EG}$ for unresolved blazars is an input to our calculation. 
Unfortunately, the exact amplitude of the angular power spectrum of the blazar component is dependent on the blazar luminosity function, and it is therefore nontrivial to obtain a self-consistent set of $I_{0}$ and blazar $C_{\ell=100}^{\rm EG}$. The angular power spectrum of blazars has been calculated by \citet{Ando2007}, based on the luminosity functions of \citet[][hereafter, \citetalias{NT06}]{NT06}. 
However, the slope of the flux distribution (log$N$-log$S$) at low fluxes for \emph{Fermi} blazars appears to be shallower than the NT06 predictions, implying a higher anisotropy than the Ando et al.\ (2007) prediction for \emph{Fermi}. In addition, 
 the level of the \emph{Fermi} IGRB has not been appreciably reduced (at least for energies $\leq 1$ GeV) compared to the determination by Strong et al.\ (2004) for EGRET data.  We therefore use the (higher) value of $C_{\ell=100}^{\rm EG}$ calculated by Ando et al.\ (2007) for the EGRET diffuse emission as our reference value.
However, we stress that an accurate, self-consistent determination of the blazar cumulative unresolved emission amplitude and anisotropy requires a detailed calculation based on luminosity functions derived from \emph{Fermi} data. While the techniques discussed here are robust regardless of the value of $C_{\ell=100}^{\rm EG}$, the sensitivity curves derived could shift up or down depending on how this value differs from our assumptions.

\subsection{Spectral and anisotropy properties of dark matter}
Properties of the dark matter particle $\chi$ such as its mass $m_{\chi}$, pair-annihilation cross-section $\langle \sigma v \rangle$, and dominant annihilation channels, in conjunction with the properties of the dark matter subhalo population, will determine the shape, amplitude, and location of a dark matter feature in the anisotropy energy spectrum of the IGRB\@.  The differential photon intensity $I$ in a direction $\psi$ from dark matter annihilation is
\begin{equation}
\label{idm}
I(\psi)=\frac{1}{4\pi} \frac{dN}{dE} \frac{\langle \sigma v \rangle}{2 m_{\chi}^{2}} \int_{\rm los} \rho^{2}(s) ds
\end{equation}
where $dN/dE$ is the photon spectrum per annihilation, and $\rho(s)$ is the dark matter density in the line-of-sight direction $\psi$ at a distance $s$.  The ``astrophysical'' factor, which encodes the mass distribution of the dark matter in subhalos and is given above by the line-of-sight integral of the dark matter density squared, is set by the structural properties of the subhalo population.

\begin{figure}[htp]
\includegraphics[width=0.47\textwidth]{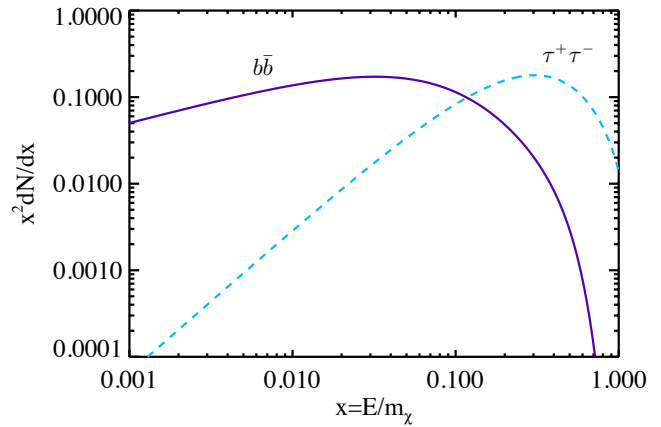}
 \caption{Continuum photon spectra for a $m_{\chi}=500$ GeV dark matter particle annihilating to $b\bar{b}$ ({\it purple solid line}) and to $\tau^+\tau^-$ ({\it blue dashed line}). \label{fig:dmspectra}} 
\end{figure}

We consider the continuum photon spectra for two benchmark annihilation channels: $\chi\chi \rightarrow b\bar{b}$, which produces a relatively soft spectrum, and $\chi\chi \rightarrow \tau^{+}\tau^{-}$, which produces a harder spectrum.  We parameterize the intensity energy spectra for these channels using the analytic approximations of \citet{fornengo_pieri_scopel_04} and show them for comparison in Figure~\ref{fig:dmspectra}.  Although we defer studying the case of gamma-ray lines ($\chi\chi \rightarrow \gamma\gamma$ or $\chi\chi \rightarrow Z^{0}\gamma$) to future work, we note that the anisotropy energy spectrum may be highly sensitive to these annihilation channels since the line feature in the intensity spectrum could lead to a dramatic variation in $f_{\mbox{\tiny{DM}}}(E)$ against the collective emission spectrum of astrophysical sources \citep[see][]{zhang_beacom_04}.

We adopt the fiducial model A1 of \citet{Ando2009} to self-consistently determine the amplitude of the angular power spectrum and the astrophysical factor entering in the normalization of the dark matter intensity.  Model A1 extrapolates the results of recent numerical simulations to a minimum subhalo mass $M_{\rm min}=10^{-6}$ $M_{\odot}$, using a mass function slope $\alpha=1.9$, which leads to a fraction $f=0.2$ of the total halo mass in substructure for this choice of $M_{\rm min}$ and normalization of the mass function.

\section{Detectability of a Dark Matter Annihilation Feature}
\label{sec:detectability}

Having defined the models for blazars and dark matter which will comprise the two-component scenarios considered in this study, we now determine in what cases the presence of the dark matter component can produce a detectable modulation in the anisotropy energy spectrum.

\subsection{Detectability Criterion}
\label{sec:criteria}

In order to quantitatively assess the sensitivity of the anisotropy energy spectrum as a tool for indirect dark matter detection, we need to define a uniform detectability criterion. We do this by testing the hypothesis that, in each scenario we examine, the anisotropy energy spectrum at energies $\geq 0.5$ GeV is consistent with a constant value, equal to the weighted average $\mu$ of all energy bins in which the measured $C_{\ell}$ is at least $2\mbox{-}\sigma$ from zero. If this hypothesis is rejected by a  $\chi^2$ test at a 95\% confidence level (CL), then we consider the particular dark matter model detectable.

To evaluate the sensitivity of this technique using \emph{Fermi} data, we assume that a fraction of the sky $f_{\rm sky}$ of 0.75 will yield data usable for this analysis, since we expect that contamination by Galactic foregrounds will be substantial in the inner Galaxy.  We estimate the number of signal (dark matter plus blazar) photons $N_{\rm s}$ that will be detected in the usable fraction of the sky in one or five years of all-sky observation time.  \emph{Fermi} operates primarily in sky-scanning mode, resulting in fairly uniform exposure across the sky; hence, we assume uniform exposure.  We take the field of view $\Omega=2.4$ sr, and approximate the effective area $A_{\rm eff}$ as a function of energy according to the reported \emph{Fermi}-LAT performance\footnote{\tt http://www-glast.slac.stanford.edu/software/IS/\\glast\_lat\_performance.htm}.

Bins in energy are logarithmic and correspond to an increase in energy of a factor of 2, with the additional constraint that at least $2\times 10^4$  signal (dark matter plus blazar) photons are detected in each bin. This is estimated by convolving the total intensity spectra of the dark matter and blazar model being tested with the effective area of \emph{Fermi} then multiplying by the observation time. The $1\mbox{-}\sigma$ statistical uncertainties in each energy bin are given by \citep[see, e.g.,][]{Knox1995}
\begin{equation}
\label{errors}
\Delta C_{\ell}^{\rm s} = \sqrt{\frac{2}{(2\ell + 1)\,\Delta\ell\, f_{\rm sky}}} \left(C_{\ell}^{\rm s} + \frac{C_{\rm N}}{W_{\ell}^{2}}\right),
\end{equation}
where $C_{\ell}^{\rm s}$ is the angular power spectrum of the signal (in our case, $C_{\ell}^{\rm tot}$) and $W_{\ell}  \! = \!  \exp(-\ell^{2}\sigma_{\rm b}^{2}/2)$ is the window function of a Gaussian beam of width $\sigma_{\rm b}$, where  $\sigma_{\rm b}$ is the energy-dependent LAT angular resolution, which we approximate by a parameterization of the reported LAT performance.  We use multipole bins of $\Delta\ell = 10$.  The noise power spectrum $C_{\rm N} = (4 \pi f_{\rm sky}/N_{\rm s})(1 + (N_{\rm b}/N_{\rm s}))$ is the sum of the Poisson noise of the signal (here blazars and dark matter) and the background (e.g., diffuse emission from Galactic cosmic ray interactions and instrument noise). The ratio of the number of background photons to the number of signal photons $N_{\rm b}/N_{\rm s}$ is taken to be 5 \citep[see, e.g.,][]{LATDiffuse2009}.

For each scenario, defined by a specific blazar and dark matter model, we test the consistency of the observable anisotropy energy spectrum with a constant value by applying a $\chi^2$ test.  The standard $\chi^2$ statistic is:
\begin{equation}\label{chi2}
\chi^2 = \sum \frac{\left(O - \mu \right)^2}{\sigma^2},
\end{equation}
where $O$ is the measured value of $C_{\ell}$ at each energy bin and $\sigma=\Delta C_\ell^{\rm s}$ is the 1-$\sigma$ error on each point\footnote{The uncertainties used in calculating the weighted average of all bins are calculated using the value of $C_\ell$ measured in each bin in Equation~(\ref{errors}). However, the uncertainties that enter Equation~(\ref{chi2}) are calculated by using the weighted average value of $C_\ell$ in Equation~(\ref{errors}) for all bins, as this is now the hypothesis that is being tested. This choice, in combination with the fact that for $\ell=100$ the probability distribution of $C_\ell$ asymptotically approaches a normal distribution, renders the uncertainties approximately Gaussian, and the CLs of the standard $\chi^{2}$ analysis apply.}.  We have attempted to appropriately estimate the measurement uncertainties taking into account the expected signals, backgrounds, and instrumental capabilities; however, we caution that underestimating the measurement uncertainties would result in an overestimate of the sensitivity of the technique and could lead to spurious detections.  To apply this technique to LAT data, a careful determination of the measurement uncertainties will be essential.

In summary, we first compute the total intensity energy spectrum of the blazar model combined with the dark matter model being tested as would be seen by \emph{Fermi} by using the spectra defined in \S Section~\ref{sec:models} and the \emph{Fermi} observational parameters described in this section. Using this intensity energy spectrum and the $C_{\ell}$ values for blazars and dark matter discussed in \S Section~\ref{sec:models}, we compute the anisotropy energy spectrum via Equation~(\ref{eq:clsum}) to determine the $C_{\ell}$ value which would be observed in each energy bin, with errors given by Equation~(\ref{errors}). Using Equation~(\ref{chi2}), we perform a $\chi^2$ test on this spectrum to determine if the anisotropy energy spectrum is consistent with a constant value at the 95\% CL. If it is not consistent with a constant, then we declare the modulation in the anisotropy energy spectrum due to the dark matter component to be detectable.

\subsection{Sensitivity to dark matter models}

\begin{figure}[t]
\includegraphics[width=0.47\textwidth]
{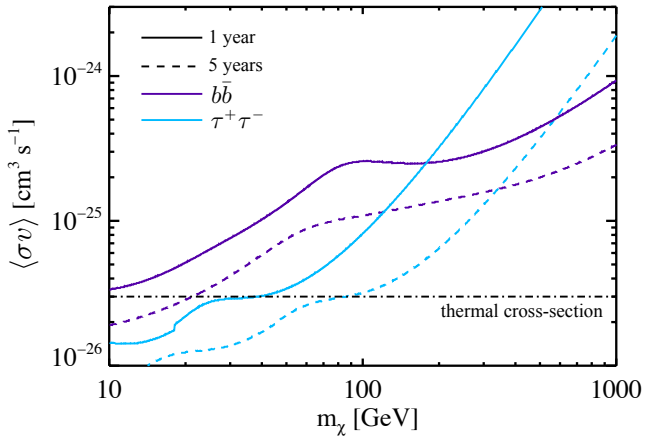}
\caption{Sensitivity of \emph{Fermi} to modulations in the anisotropy energy spectrum due to dark matter annihilation in substructure for the test criterion specified in Section~\ref{sec:criteria}.  The minimum detectable annihilation cross section to $b\bar{b}$ ({\it purple}) or $\tau^{+}\tau^{-}$ ({\it light blue}) is shown assuming one year ({\it solid}) or five years ({\it dashed}) of observations.  Dark matter models above the curves produce a feature detectable at 95\% CL.\label{fig:mchisigv}}
\end{figure}

We now examine the sensitivity of this technique to dark matter particle properties.  In Figure~\ref{fig:mchisigv}, we show the detectable region of the $m_{\chi}$-$\langle \sigma v \rangle$ parameter space for one and five years of \emph{Fermi} observations assuming our reference blazar model and annihilation to $b\bar{b}$ or $\tau^{+}\tau^{-}$.  Models above the lines for each annihilation channel are detectable for the test criterion defined in Section~\ref{sec:criteria}.  Using this criterion, we find that annihilation cross-sections of order $3 \times 10^{-26}$ cm$^{3}$ s$^{-1}$, the expected value for a thermal relic, are within the reach of \emph{Fermi} for the $\tau^{+}\tau^{-}$ channel for masses up to $\sim 100$ GeV (50 GeV) with five years (one year) of observations, and cross sections within a factor of 10 of thermal are detectable for a large range of masses (between $\sim 10$ and $\sim 400$ GeV) within five years for this annihilation channel.  For annihilation to $b\bar{b}$, cross sections within a factor of 10 of thermal are detectable in five years for the entire range of masses considered here, $10$ GeV $< m_{\chi} < 1$ TeV.  It is notable that this technique is sensitive to a viable region of the parameter space; existing constraints robustly exclude total annihilation cross sections larger than $\sim 10^{-23}$ cm$^{3}$ s$^{-1}$ for the mass range considered here \citep{beacom_bell_mack_07}.  Reionization constraints from the observed optical depth of the universe more strongly limit the cross-section for these annihilation channels to $\langle \sigma v \rangle \lesssim 10^{-24}$ cm$^{3}$ s$^{-1}$ for $m_{\chi} \lesssim 100$ GeV \citep[e.g.,][]{slatyer_padmanabhan_finkbeiner_09, huetsi_hektor_raidal_09, cirelli_iocco_panci_09}.  

The subtle bump-like features in Figure~\ref{fig:mchisigv}, present for both annihilation channels, are a result of the specific criterion we chose for the statistical test, and not an indication of a rapid change in the properties of the dark matter model at those masses.  Specifically, the energy range and multipole we selected for this analysis lead to weaker sensitivity in these regions.  This can be understood in the $b\bar{b}$ one-year case by noting that for $100 \lesssim m_{\chi} \lesssim 200$ GeV, this channel produces a large number of photons at energies of a few GeV, where \emph{Fermi} can generally make a precise measurement of the anisotropy.  As the blazar and dark matter intensity spectra fall above this energy, the number of energy bins in which a precise measurement of the angular power spectrum can be made is limited due to low photon statistics.  In this situation, a deviation in the anisotropy energy spectrum (i.e., the transition at high energies to the anisotropy being dominated by blazars) is difficult to detect by our criterion using a small number of points which may have relatively large measurement uncertainties.  The features in the $\tau^{+}\tau^{-}$ sensitivity curves are present for similar reasons.  Alternative parameters for the statistical test (e.g., a different multipole or energy range) would modify these features.  The sensitivity of this technique to these models may improve by optimizing the detection criterion of the test considered here, or by the application of more tailored likelihood analysis methods to test these scenarios \citep[e.g.,][]{dodelson_belikov_hooper_etal_09}; however, such a detailed  study is beyond the scope of this work.

\begin{figure}[t]
\includegraphics[width=0.47\textwidth]
{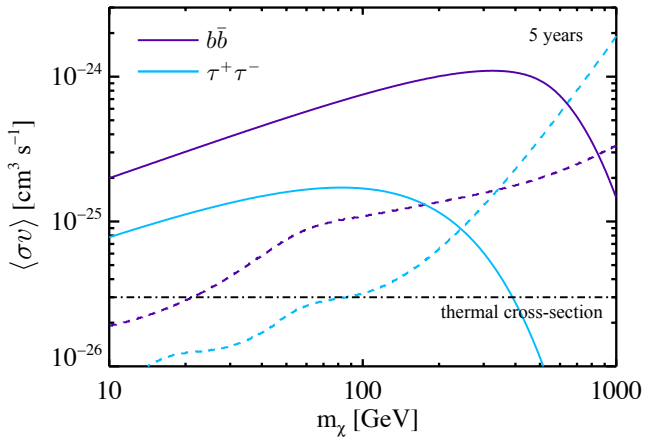}
\caption{Annihilation cross-section $\langle \sigma v \rangle$ as a function of dark matter mass $m_{\chi}$ above which the dark matter intensity exceeds the intensity of the reference blazar model at any energy above 0.5 GeV ({\it solid curves}), for annihilation to $b\bar{b}$ ({\it purple}) and $\tau^{+}\tau^{-}$ ({\it light blue}).  Sensitivity curves from Figure~\ref{fig:mchisigv} for five years of observation assuming the reference blazar model are shown for comparison ({\it dashed curves}).  The dark matter signal is subdominant in the intensity energy spectrum at energies above 0.5 GeV for models below the solid curves.\label{fig:dmdom}}
\end{figure}

\begin{figure}[t]
\includegraphics[width=0.47\textwidth]
{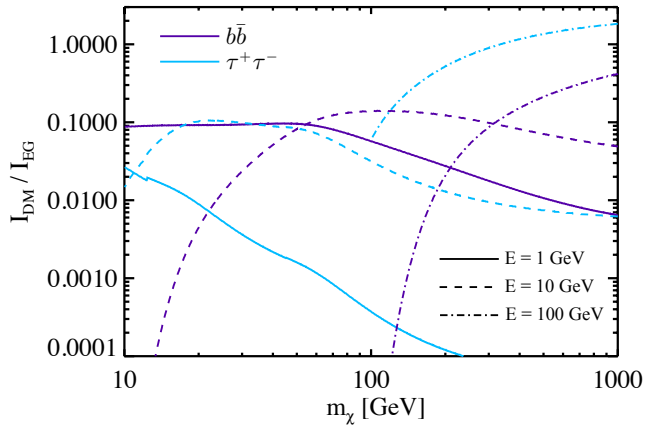}
\caption{Ratio of dark matter to blazar intensity $I_{\rm DM}/I_{\rm EG}$ at three reference energies as a function of dark matter mass $m_{\chi}$ for detectable models (annihilation cross sections) corresponding to the five-year sensitivity curves shown in Figure~\ref{fig:mchisigv}.  Relative intensity of the dark matter signal is shown at $E = 1$, 10, and 100 GeV (as labeled) for annihilation to $b\bar{b}$ ({\it purple}) and $\tau^{+}\tau^{-}$ ({\it light blue}).  The $E=100$ GeV curves do not extend to masses below 100 GeV because no dark matter annihilation photons are produced at 100 GeV for $m_{\chi} < 100$ GeV.\label{fig:relintens}}
\end{figure}

As an indication of the relative level of dark matter contribution to the IGRB in our models, we show in Figure~\ref{fig:dmdom} the annihilation cross section $\langle \sigma v \rangle$ at which the dark matter intensity exceeds that of the reference blazar model at any energy $E>0.5$ GeV as a function of dark matter mass $m_{\chi}$ for the two annihilation channels considered ({\it solid curves}).  As expected, for large dark matter masses ($m_{\chi} \gtrsim$ 300 GeV), the intensity of the reference blazar model is exceeded even at fairly small annihilation cross sections for the $\tau^{+}\tau^{-}$ channel, due to the exponential cut-off of the blazar intensity above a few tens of GeV and the relatively hard continuum spectrum of this channel.  Annihilation to $b\bar{b}$ produces a softer continuum spectrum, and consequently larger annihilation cross sections are required to exceed the blazar intensity for the mass range considered here than in the $\tau^{+}\tau^{-}$ case.  The detectable region of the parameter space from Figure~\ref{fig:mchisigv} for five years of observation is shown for comparison ({\it dashed curves}).  Comparing the regions indicated by the solid and dashed curves, it is clear that the dark matter component does not need to be dominant in the intensity to produce a signal detectable by this technique.

In Figure~\ref{fig:relintens}, the ratio of the dark matter to blazar intensity at 1, 10, and 100 GeV is shown for dark matter models detectable within five years at 95\% CL\@.  The normalization of the dark matter intensity at each mass is set by the annihilation cross section corresponding to the five-year sensitivity curves shown in Figure~\ref{fig:mchisigv} for each annihilation channel.  A dark matter intensity $\lesssim 10$\% of the blazar intensity at $E=1$ and 10 GeV is sufficient to produce a detectable feature in the anisotropy energy spectrum for the mass range considered here (10 GeV - 1 TeV).  At $E=100$ GeV, detectable models contribute a larger intensity relative to the blazar intensity, up to $\sim 40$\% at $m_{\chi}=1$ TeV for annihilation to $b\bar{b}$ and up to $\sim 200$\% at $m_{\chi}=1$ TeV for annihilation to $\tau^{+}\tau^{-}$.  However, we note that our reference blazar model does not account for most of the \emph{Fermi} measurement of the IGRB at $E=100$ GeV (see Figure~\ref{fig:diffuse}, assuming the power-law IGRB can be extended to $E=100$ GeV), so dark matter models that exceed our blazar model intensity at this energy by a factor of $\sim 2$ are still acceptable given the measured IGRB level.

\subsection{Dependence on blazar model parameters}
\label{sec:blazardep}
\begin{figure}[htp]
\includegraphics[width=0.48\textwidth]{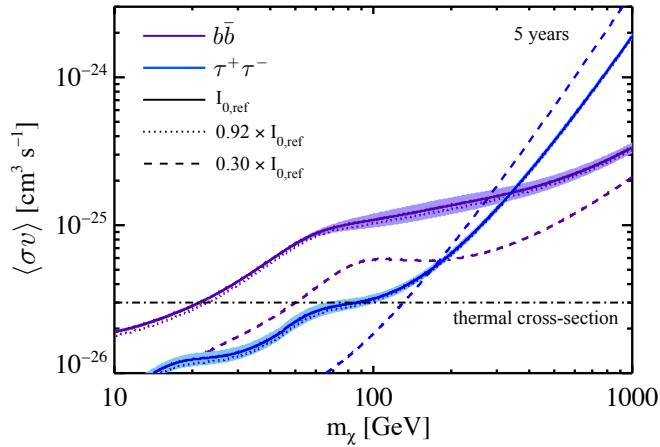}
\caption{Minimum value of the annihilation cross section for which a dark matter annihilation signal is detectable at 95\% CL in five years of observation with \emph{Fermi} as a function of $m_{\chi}$, for annihilation to $b\bar{b}$ ({\it purple}) or $\tau^{+}\tau^{-}$ ({\it blue}).  The height of the band at each dark matter mass represents the uncertainty in detectability entering through the $1\mbox{-}\sigma$ uncertainties in the blazar spectral parameters $\alpha_0$ and $\sigma_0$, assuming the reference blazar intensity normalization; the solid line within each band shows the sensitivity obtained for the maximum likelihood values of the spectral parameters.  The sensitivity curves for the blazar intensity normalized to 92\% ({\it dotted lines}) and 30\% ({\it dashed lines}) of the reference model normalization, assuming the maximum likelihood spectral parameters, are shown for comparison.\label{fig:blazarsens}}
\end{figure}

We now examine how the detectability of a dark matter feature in the anisotropy energy spectrum depends on the overall normalization and the spectral shape of the blazar component of the IGRB\@.  We show in Figure~\ref{fig:blazarsens} the sensitivity of this technique to variations within the $1\mbox{-}\sigma$ uncertainty in our determination of the blazar spectral parameters $\alpha_0$ and $\sigma_0$ ({\it shaded bands}) and variations in the blazar normalization, assuming five years of observations.  The sensitivity for the blazar intensity normalized to 92\% ({\it dotted lines}) and 30\% ({\it dashed lines}) of the reference model normalization, assuming the maximum likelihood spectral parameters, is also shown.  These results indicate that the uncertainty in the test sensitivity introduced by uncertainties in $\alpha_0$ and $\sigma_0$ is small compared with the range of detectable signals.  Reducing the blazar normalization to the extent shown here improves our test sensitivity slightly, although it is important to keep in mind that our simple 2-component model would become unrealistic for very small blazar normalizations, since if blazars are a subdominant component of the measured IGRB we expect other source classes, not considered here, to contribute significantly.  Other scenarios for astrophysical sources of the IGRB will be addressed in a future publication. We note, however, that in the likely scenario where a population of numerous but individually faint sources  (such as star-forming galaxies or quiescent blazars) contributes a substantial fraction of the IGRB, the angular power at $\ell \sim 100$ could still be dominated by bright blazars, since such an additional population would appear very isotropic in comparison on small scales \citep[e.g.,][]{ando_pavlidou_09}. In this case, the additional component would behave as an almost isotropic ``background'' on top of our bright blazar plus dark matter ``signal''.

\section{Discussion}\label{sec:discussion}

We have examined the potential of the anisotropy energy spectrum of the IGRB to identify a dark matter annihilation signal for the case that the main contributors to the IGRB are unresolved blazars and Galactic dark matter subhalos. We have evaluated the detectability of a dark matter signature as a function of the properties of the dark matter particle and of the blazar population using a model-independent detectability criterion, namely the consistency of the anisotropy energy spectrum with a constant value over varying energies. 

We find that modulations in the anisotropy energy spectrum are a sensitive probe of dark matter models.  Using our reference blazar model and assuming annihilation to $\tau^+\tau^-$, \emph{Fermi} could detect in one year a feature from dark matter for annihilation cross sections as small as that expected for a thermal relic for particle masses below $\sim 50$ GeV.  In five years, cross sections within a factor of 10 of thermal produce a detectable feature for $10 \lesssim m_{\chi} \lesssim 1$ TeV for annihilation to $b\bar{b}$, and $10 \lesssim m_{\chi} \lesssim 400$ GeV for annihilation to $\tau^+\tau^-$.  The dependence of the test sensitivity on the blazar model parameters (spectral shape and overall normalization) is minimal and does not affect our conclusions qualitatively.
 
We have treated the simple case where only blazars and Galactic dark matter have a significant contribution to the IGRB\@. In reality, other classes of sources, such as extragalactic dark matter, normal star-forming and starburst galaxies, and galaxy clusters may also be significant IGRB components. We plan to study the effect of these additional contributions on Galactic dark matter detectability in a future publication. 

Recent results from \emph{Fermi} \citep{Ajello10} indicate that the total contribution of sources of the same classes as those comprising the majority of \emph{Fermi}-resolved sources (such as bright blazars) may comprise as little as 10\%-30\% of the total IGRB intensity. In that case, the additional IGRB photons must originate in truly diffuse processes or unresolved objects that are underrepresented in the set of already-resolved \emph{Fermi} sources. Examples of the latter are quiescent blazars \citep[e.g.,][]{stecker-1996-464}, star-forming galaxies \citep{FPP10}, and dark matter annihilation in Galactic or extragalactic structures. However, as discussed in \S Section 3.3, it is likely that even if the intensity is not dominated by bright blazars, the angular power at small scales is: numerous but individually faint sources will contribute very little to the total anisotropy. 

There is at least one guaranteed contribution to the IGRB which is expected to be similar in spectral shape with dark matter annihilating to $b\bar{b}$ (although peaking at lower energies than most plausible dark matter candidates): millisecond pulsars, which constitute a known \emph{Fermi} source population with resolved members \citep{abdo_others_09}, and which have a spatial distribution  extending to high Galactic latitudes \citep[see, e.g.,][]{faucher-giguere_loeb_10}. However, their contribution would be limited to the lower end of the energy range we have considered, as their spectra generally cut off exponentially above $\sim$ few GeV. In addition, since the MSPs are correlated with the Galactic plane, the large-scale morphology of their contribution to the diffuse emission would show a gradient away from the plane, which could be used to identify its origin. Finally, the spectral signature of millisecond pulsars is fairly well constrained, so in a more detailed study the modulation of the anisotropy energy spectrum due to the presence of millisecond pulsars can be explicitly modeled and taken into account. 

The sensitivity estimated in this work is based on a generic test for the presence of any detectable deviation from energy invariance in the anisotropy energy spectrum; the results obtained depend on the parameters defining the statistical test, such as the choice of multipole and energy bins.  To straightforwardly define our statistical test, we have not fully optimized the parameters of the test for each set of dark matter parameters.  This is reflected in the features in the sensitivity curves shown in Figure~\ref{fig:mchisigv}, which arise from the parameters chosen for our statistical test, and can change qualitatively for different test parameters.  We expect that instead testing the likelihood of a specific dark matter and blazar model producing observed data will extend the sensitivity of this technique, and defer that analysis to future work.

The anisotropy and overall intensity in the dark matter subhalo gamma-ray signal was calculated in a manner as self-consistent as possible based on one of the fiducial models of \citet{Ando2009}. However, this model does not take into account the radial dependence of subhalo concentrations \citep[e.g.,][]{kuhlen_diemand_madau_08} which may result in increased anisotropy and intensity from subhalos. In addition, a steeper subhalo mass--luminosity relation than employed in this model, especially for subhalos below $\sim 10^{6}$ $M_{\odot}$, is predicted from the simulations of \citet{kuhlen_diemand_madau_08}.  Using this steeper relation would decrease the contribution from low-mass subhalos to the intensity, which would likely increase the overall amplitude of the angular power spectrum from Galactic substructure.  The anisotropy of the blazar signal was calculated based on the luminosity functions of \citetalias{NT06}. However, first results from \emph{Fermi} \citep{Ajello10} may indicate that the number density of blazars is lower than the predictions of \citetalias{NT06}, which, if confirmed, would imply that the anisotropy in the blazar signal may also be higher than the values considered here.  These effects would likely lead to improved detectability and increased constraining power since we are generally in a low-statistics regime, and thus larger amplitude anisotropies can be measured more easily.

\acknowledgements
{We are grateful to Shin'ichiro Ando, Alessandro Cuoco, and Tonia Venters for helpful discussions, and to John Beacom and Anthony Readhead for valuable feedback.  B.S.H. thanks CCAPP, and J.S.-G. and V.P. thank the Pauli Center for Theoretical Studies at the University of Z\"{u}rich, for hospitality during the completion of this study.  B.S.H. also acknowledges the Summer Undergraduate Research Fellowship (SURF) program at Caltech, and in particular Barbara and Stanley Rawn, Jr. and the Alain Porter Memorial SURF Fellowship for their support.  J.S.-G. was supported in part by NSF CAREER Grant PHY-0547102 (to John Beacom).  V.P. acknowledges support for this work provided by NASA through Einstein Postdoctoral Fellowship grant number PF8-90060 awarded by the Chandra X-ray Center, which is operated by the Smithsonian Astrophysical Observatory for NASA under contract NAS8-03060.  This work was partially supported by NASA through the Fermi GI Program grant number  NNX09AT74G\@.}

\bibliography{aniso_sens_v2}

\end{document}